\documentclass[paper]{JHEP3} 

\usepackage{epsfig,multicol,multirow}
\usepackage{amsmath,subfigure,latexsym,amssymb,cite}
\usepackage{here}

\newcommand\fverb{\setbox\fverbbox=\hbox\bgroup\verb}
\newcommand\fverbdo{\egroup\medskip\noindent%
			\fbox{\unhbox\fverbbox}\ }
\newcommand\fverbit{\egroup\item[\fbox{\unhbox\fverbbox}]}
\newbox\fverbbox

\newcommand {\beq} {\begin{equation}}
\newcommand {\eeq} {\end{equation}}
\newcommand {\beqa}{\begin{eqnarray}}
\newcommand {\eeqa}{\end{eqnarray}}

\newcommand {\tr}{{\rm tr\,}}

\newcommand {\n} {\nonumber}
\newcommand{\1}{\mbox{1}\hspace{-0.25em}\mbox{l}}

\allowdisplaybreaks

\title{Local field theory
from the expanding universe at late times
in the IIB matrix model}

\author{Jun Nishimura${}^{ab}$ and Asato Tsuchiya${}^{c}$
\vspace*{0.5cm} \\
\llap{$^a$}Department of Particle and Nuclear Physics,\\
Graduate University for Advanced Studies (SOKENDAI),\\
Tsukuba, Ibaraki 305-0801, Japan\\
\llap{$^b$}High Energy Accelerator Research Organization (KEK),\\
Tsukuba, Ibaraki 305-0801, Japan\\
\llap{$^c$}Department of Physics, Shizuoka University,\\
836 Ohya, Suruga-ku, Shizuoka 422-8529, Japan
\vspace*{0.5cm} \\
\email{jnishi@post.kek.jp, satsuch@ipc.shizuoka.ac.jp}}

\preprint{KEK-TH-1565}

\abstract{
Recently we have shown
that (3+1)-dimensional expanding universe 
appears dynamically and uniquely from the Lorentzian version 
of the IIB matrix model,
which is considered as
a nonperturbative formulation of superstring theory.
Similarly, it is possible that the Standard Model appears uniquely
at the electroweak scale from the same model at late times.
In order to pursue such a possibility,
we discuss how to
derive the effective local field theory for the massless modes 
that appear at late times in the same formulation.
As a concrete example, we consider the massless modes associated
with the spontaneous breaking of
(9+1)-dimensional 
Poincare symmetry and supersymmetry.
}


\begin{document}

\section{Introduction}
\label{sec:introduction}

The use of a nonperturbative formulation
is often indispensable for understanding 
important dynamical properties
of a theory. For instance, confinement of quarks in 
quantum chromodynamics can be most vividly understood
by using the lattice gauge theory \cite{Wilson:1974sk}.
This might also be the case in string theory.
Since 1980s, it 
has been
considered that superstring theory has
infinitely many vacua that are perturbatively stable.
Each of them corresponds to space-time of various dimensionality
with various kinds of matter and gauge symmetry.
However, it is, of course, possible that the theory has
a unique nonperturbative vacuum, which represents
a (3+1)-dimensional space-time with the Standard Model (SM)
particles
propagating on it 
at low energy.

In order to pursue this possibility,
we consider the IIB matrix model \cite{IKKT},
which is 
a nonperturbative formulation of superstring theory 
based on
type IIB string theory in (9+1) dimensions.
This model is particularly suitable 
for addressing the above issue
since the space-time as well as the fields on it
is treated as matrix degrees of freedom in a unified fashion.
In ref.~\cite{Kim:2011cr}
we have found by Monte Carlo simulation
that time as well as space emerges dynamically
in the Lorentzian version of the 
IIB matrix model \cite{Kim:2011cr}.
Furthermore, 
we were able to extract the time-evolution,
which turned out to 
be surprising:
9-dimensional space undergoes 
spontaneous breaking of rotational symmetry 
at some critical time, after which
only three directions start to expand.
%

This result strongly suggests that
(3+1)-dimensional expanding universe appears dynamically
from the Lorentzian model.
It should be emphasized that the space-time dimensionality
seems to be uniquely determined as a result of 
nonperturbative dynamics.
Likewise 
the SM may emerge
uniquely from the IIB matrix model.
%
%

The mechanism of the spontaneous symmetry breaking (SSB) 
relies crucially on 
noncommutativity of 
the space-time represented by 10 bosonic matrices.
Therefore, 
an important issue
is whether the usual commutative space-time appears at later times.
We addressed this issue by studying the classical equations
of motion, which
are expected to be valid
at late times \cite{Kim:2011ts,Kim:2012mw}.
There are actually infinitely many solutions representing
commutative (3+1)-dimensional space-time.
Moreover, we found a simple solution 
with an expanding behavior, which
naturally solves the cosmological constant problem \cite{Kim:2012mw}.
We consider that there exists a unique solution of this kind
that dominates
the partition function of the matrix model
at late times.


In this paper we discuss how a local field theory appears
at late times in the same model.
In fact it has been considered 
rather difficult to obtain a local field theory
from the IIB matrix model in the \emph{Euclidean} case \cite{Iso:1999xs}.
This has something to do with the absence of a mechanism
that makes the classical approximation valid
in performing the 
matrix integration \cite{Hotta:1998en,Ambjorn:2000bf}.
In the present \emph{Lorentzian} case, however, the expansion 
of the Universe makes the classical approximation to the 
matrix integration more and more valid at late times.
This implies that the quantum fluctuations around a classical solution 
are strongly suppressed.
In such a situation,
a local field theory is naturally obtained as an effective
theory for the massless fluctuations around a classical solution
representing a commutative space-time as we will discuss later.
%
%
We demonstrate our idea by
considering the Nambu-Goldstone (NG) modes and their extension
as a concrete example of the massless modes.
These NG modes are the ones that are associated with the
SSB of (9+1)-dimensional Poincare symmetry and supersymmetry (SUSY)
due to the chosen classical solution.


Once we identify
the classical solution (representing a commutative space-time)
that dominates at late times,
we can uniquely derive the 
local field theory for
the massless modes based on our idea.
This local field theory 
is considered to be the effective theory
at a scale a few orders of magnitudes
lower than the Planck scale,
where the quantum gravitational effects 
and the effects of stringy excited states become negligible.
By studying the dynamics of this effective theory using usual
field theoretical methods, one should be able to see whether
the SM emerges at the electroweak scale.

The rest of this paper is organized as follows.
In section \ref{sec:IIBmm} 
we briefly review the IIB matrix model.
In section \ref{sec:local-fields}
we discuss how a local field theory appears in the Lorentzian 
model as an effective theory for the massless modes.
%
In section \ref{sec:NG-modes} we demonstrate our idea
by considering the NG modes 
as a concrete example of the massless modes.
Section \ref{sec:concl} is devoted to a summary and discussions.

\section{Matrix model for nonperturbative superstring theory}
\label{sec:IIBmm}

Let us briefly review the IIB matrix model \cite{IKKT},
which is a nonperturbative formulation of superstring theory.
The action is given by
$S = S_{\rm b} + S_{\rm f}$ with
%
\beqa
S_{\rm b} &=& -\frac{1}{4g^2} \, \tr \Bigl( [A_{M},A_{N}]
[A^{M},A^{N}] \Bigr) \ , \n  \\
S_{\rm f}  &=& - \frac{1}{2g^2} \,
\tr \Bigl( \Psi _\alpha (\, {\cal C} \,  \Gamma^{M})_{\alpha\beta}
[A_{M},\Psi _\beta] \Bigr)  \ ,
\label{action}
\eeqa
where $A_M$ ($M = 0,\cdots, 9$) and
$\Psi_\alpha$ ($\alpha = 1,\cdots , 16$) are
$N \times N$ 
Hermitian matrices.
The Lorentz indices $M$ and $N$ are
contracted
using the metric
$\eta={\rm diag}(-1 , 1 , \cdots , 1)$.
The $16 \times 16$ matrices $\Gamma ^M$ are
10D
gamma matrices after the Weyl projection,
and the unitary matrix ${\cal C}$ is the
charge conjugation matrix.
The action has manifest SO(9,1) symmetry,
under which $A_{M}$ and $\Psi _\alpha$ transform as a
vector and
a Majorana-Weyl spinor, respectively:
\begin{align}
\left\{ \begin{array}{ll}
 \delta_{\rm L} A_{M} &= \epsilon_{MN}A^{N} \ , \\
\delta_{\rm L} \Psi &=\frac{1}{2}\epsilon_{MN}\Gamma^{MN}\Psi \ .
\end{array} \right.
\label{Lorentz transformation}
\end{align}
The model also has the fermionic symmetries
\begin{align}
& \left\{ \begin{array}{ll}
\delta^{(1)}A_{M} &=i\epsilon_1 {\cal C}\Gamma_{M} \Psi  \ , \\
\delta^{(1)}\Psi &=\frac{i}{2}\Gamma^{MN}[A_{M},A_{N}]\epsilon_1 \ , 
\end{array} \right.
\label{susy1} \\
& \left\{ \begin{array}{ll}
\delta^{(2)}A_{M} &=0 \ , \\
\delta^{(2)}\Psi &=\epsilon_2 \1_N \ ,
\end{array} \right.
\label{susy2}
\end{align}
as well as the bosonic symmetry
\begin{align}
\left\{ \begin{array}{ll}
\delta_{\rm T} A_{M} &=c_{M} \1_N  \ , \\
\delta_{\rm T} \Psi &=0 \ ,
\end{array} \right.
\label{translation}
\end{align}
where $\1_N$ is the $N\times N$ unit matrix.
Let us denote the generators of (\ref{susy1}), (\ref{susy2}) and
(\ref{translation}) 
by $Q^{(1)}$, $Q^{(2)}$ and
$P_{M}$, respectively, and define
$\tilde{Q}^{(1)}=Q^{(1)}+Q^{(2)}$,
$\tilde{Q}^{(2)}=i(Q^{(1)}-Q^{(2)})$.
Then, one obtains the algebra
\begin{align}
[\epsilon_1{\cal C}\tilde{Q}^{(i)},\epsilon_2{\cal C}\tilde{Q}^{(j)}]
=-2\delta^{ij}\epsilon_1{\cal C}\Gamma^{M}\epsilon_2P_{M}
\label{N2SUSY}
\end{align}
with $i,\;j=1,2$, which is nothing but the ten-dimensional ${\cal N}=2$ SUSY
if one identifies $P_{M}$ 
with the momentum.
%
Thus we find that the model possesses the maximal SUSY that type IIB
superstring theory has.

There are various evidences
that the model gives a nonperturbative formulation of 
superstring theory.
First of all, the action (\ref{action})
can be viewed as a matrix regularization of the worldsheet action
of type IIB superstring theory in a particular gauge known 
as the Schild gauge \cite{IKKT}.
It has also been argued that
configurations of block-diagonal matrices
correspond to a collection of disconnected worldsheets 
with arbitrary genus.
Therefore, instead of being 
equivalent
just to the worldsheet theory,
the large-$N$ limit of the matrix model is expected to be 
a second-quantized theory of type IIB superstrings,
which includes multi-string states.
Secondly, D-branes are represented as classical solutions in the 
matrix model, and the interaction between them calculated
at one loop reproduced correctly the known results
from type IIB superstring theory \cite{IKKT}.
Thirdly, one can derive
the light-cone string field theory for the type IIB case
from the matrix model \cite{FKKT} with a few assumptions.
In the matrix model, one can define
the Wilson loops, which can be naturally identified
with the creation and annihilation operators of strings.
Then, from the Schwinger-Dyson equations for the Wilson loops,
one can actually obtain the string field Hamiltonian. 

In all these connections to string theory,
it is crucial that the model has the symmetry (\ref{N2SUSY}).
When we identify it with 
the ten-dimensional ${\cal N}=2$ SUSY,
the symmetry (\ref{translation}) is identified
with the translational symmetry in ten dimensions,
which implies that 
the eigenvalues of $A_{M}$ should 
be identified with 
the coordinates of ten-dimensional space-time.
This identification 
is consistent with the one adopted in stating
the evidences listed in the previous paragraph,
and shall be used throughout this paper as well.

\section{Local fields from the matrix model}
\label{sec:local-fields}

An important feature of the IIB matrix model is that
space-time is treated as a part of dynamical degrees of freedom.
It is therefore possible that four-dimensional space-time appears
dynamically. The Monte Carlo results of ref.~\cite{Kim:2011cr}
demonstrate that this 
really happens.
(See ref.~\cite{Nishimura:2011xy} and references therein 
for earlier works on the Euclidean version of the model.)

The dominant configurations observed in Monte Carlo simulation \cite{Kim:2011cr}
show that the elements of $A_m$ $(m=4,\cdots,9)$ are small
compared with those of $A_{\mu}$ $(\mu=0,1,2,3)$.
Moreover, in the basis which diagonalizes $A_0$,
the matrices $A_{i}$ $(i=1,2,3)$ have a band diagonal structure,
which enables us to extract the time-evolution of the space.
Due to the finite size of the matrices, however, we were able to 
probe only the early time behaviors so far.
On the other hand, the classical solutions \cite{Kim:2012mw},
which are expected to describe the late time behaviors of 
the matrix model,
have $A_{\mu}$ with a tri-diagonal structure and $A_m=0$.

Motivated by these results, 
we consider that
a classical solution $A_M=\hat{A}_M$ and $\hat{\Psi}=0$ 
dominates at late times, 
where $\hat{A}_{\mu}$ are ``close to diagonal" and $\hat{A}_m$ are small.
In this situation, the diagonal elements $(\hat{A}_{\mu})_{ii}$ correspond
to the space-time points $(x_i)_{\mu}$ in (3+1) dimensions. 
Furthermore, we focus on a sufficiently 
small space-time region, 
and assume that
the space-time can be approximated by 
the direct product of 
$(3+1)$-dimensional Minkowski space 
and 6D small extra dimensions. 
This enables us to give a more precise definition of
being ``close to diagonal'' stated above.
By this we actually mean that the $(i,j)$
elements corresponding to $\sqrt{(x^{\mu}_i-x^{\mu}_j)^2} \gg \ell$ are small,
where $\ell$ is some length parameter
that appears dynamically.
We consider that matrices satisfying this condition represent 
commutative (3+1)-dimensional space-time.

Let us then consider the fluctuations 
around the above classical solution as
\begin{align}
A_M&=\hat{A}_M+\tilde{A}_M \ , \nonumber\\
\Psi&=\tilde{\Psi} \ .
\label{fluctuation}
\end{align}
Substituting (\ref{fluctuation}) into (\ref{action}) 
yields the action for $\tilde{A}_M$ and
$\tilde{\Psi}$. The terms linear in $\tilde{A}_M$ and $\tilde{\Psi}$ are 
absent 
due to the equations of motion.
By diagonalizing the quadratic term, we can identify
the low-lying modes.
For the classical solution we are considering,
it is expected that
the fluctuations for the low-lying modes are also
close to diagonal; \emph{i.e.}, 
\begin{align}
(\tilde{A}_M)_{ij}=0 \mbox{~and~}
(\tilde{\Psi})_{ij}=0 \quad
\mbox{for~} \sqrt{(x^{\mu}_i-x^{\mu}_j)^2} \gg \ell \ .
\label{local-fluctuation}
\end{align}
We will see that this is indeed the case for a concrete example
in section \ref{sec:lft-NG}.
For general low-lying modes, the argument is given as follows.
Note first that the action of the IIB matrix model is written 
in terms of matrix products.
This guarantees that the quadratic term for the modes 
satisfying (\ref{local-fluctuation}) is local;
namely
it only contains such terms as
$(\tilde{A}_{M})_{ij}(\tilde{A}_M)_{i'j'}$ 
or $(\tilde{\Psi})_{ij} (\tilde{\Psi})_{i'j'}$,
where $x_i$, $x_j$, $x_{i'}$ and $x_{j'}$ are close to each other.
On the other hand, the quadratic term for
those modes which do not satisfy
(\ref{local-fluctuation}) is non-local and 
becomes large. Hence those modes cannot be a low-lying mode.

The quadratic term for the low-lying modes
would become the kinetic term and the mass term,
in general, in the continuum limit 
due to the (3+1)-dimensional Poincare invariance of the background.
However, the mass term would involve a mass of the order of the
Planck scale, and hence it cannot appear for the low-lying modes.
Thus the low-lying modes naturally correspond to the massless fields
in (3+1)-dimensions.
Obviously the locality of the effective action holds also for high order
terms,
from which we can read off the local field theory for the low-lying modes.

The appearance of a local field theory from 
the Lorentzian model
is somewhat analogous to the lattice construction
of SUSY gauge theory \cite{Kaplan:2002wv},
which uses a matrix model with orbifolding conditions.
An important difference, however, is that in our case, 
the orbifolding conditions
are not imposed,
but similar effects appear dynamically.



In order to introduce the gauge symmetry in 
the effective field theory obtained from the matrix model,
we consider the background as in ref.~\cite{Iso:1999xs}
\begin{align}
\hat{A}_M'=\hat{A}_M\otimes \1_k \ , \;\;\; \hat{\Psi}'=0 \ ,
\label{extended background}
\end{align}
where $\1_k$ is the $k\times k$ unit matrix and
$\hat{A}_M$ is the background discussed above.
The matrix model is now considered with the matrix size $N'=Nk$,
where the matrix index is represented by a pair of indices $(i,a)$
with $i=1,\cdots,N$ and $a=1,\cdots,k$.
The extended background (\ref{extended background}) is 
still a classical solution, 
and it represents a situation analogous to that of 
$k$ D3-branes lying on top of each other, 
which will be discussed in section \ref{sec:ssb-dbrane}.

%
%
%

The background (\ref{extended background})
is invariant under a ${\rm U}(N')$ transformation
\begin{align}
\hat{A}_M' \rightarrow U \hat{A}_M' U^{\dagger} \ , \quad 
\mbox{where}  \quad
U_{ia,jb}=\delta_{ij}u^{(i)}_{ab}
\end{align}
with 
$u^{(i)} \in {\rm U}(k)$.
This invariance becomes the U($k$) gauge symmetry of the effective
theory.
The
${\rm U}(k)$ gauge field
is expected to appear
from 
%
$(\tilde{A}_{\mu})_{ia,jb}$. 
Since all the matter fields are 
in the adjoint representation,
the ${\rm U}(1)$ part is decoupled
and we actually obtain an ${\rm SU}(k)$ grand unified theory (GUT)
for the low-lying modes.\footnote{We may also obtain a theory
with a semi-simple gauge group by modifying the $\1_k$ part
in eq.(\ref{extended background}).}  

\section{Nambu-Goldstone modes --- as an example}
\label{sec:NG-modes}

In this section we demonstrate our idea to 
identify the local field theory 
for a concrete example.
In the Lorentzian matrix model,
the results of Monte Carlo simulation and the classical analysis
suggest that (9+1)-dimensional Poincare symmetry and 
${\cal N}=2$ SUSY
are spontaneously broken at late times.
The Nambu-Goldstone modes associated with this SSB are expected
to appear as massless modes.
Let us therefore discuss the effective field theory 
for these massless modes.
First we consider a similar issue in 
the well-known case of the D-brane background in string theory.

\subsection{Spontaneous symmetry breaking in the D-brane background}
\label{sec:ssb-dbrane}

For concreteness, we consider 
type IIB superstring theory
with a background of $k$ D3-branes lying on
top of each other in ten-dimensional flat Minkowski space.
The low energy effective theory on the worldvolume of such D-branes
is known to be four-dimensional ${\cal N}=4$ ${\rm U}(k)$ super
Yang-Mills theory,
which contains the gauge field, six adjoint scalars and 
four adjoint Weyl fermions.
The existence of the D-branes breaks continuous symmetries,
and
the associated NG modes can be identified
in the low energy effective theory
as we discuss below.

First, the translational invariance in six dimensions transverse to the D-branes
is broken spontaneously. 
The associated NG bosons
are the ${\rm U}(1)$ part of six adjoint scalars.
Secondly, half of the 32 supersymmetries of type IIB string theory are
broken since the D-brane configuration under discussion is half-BPS. 
The associated NG fermions
are the ${\rm U}(1)$ part of four Weyl fermions.
%

In fact, the ${\rm U}(k)$ gauge symmetry 
appears due to
the Chan-Paton factor at the ends of open strings.
Hence, each of the above NG modes is enhanced to a set of massless modes which
form the adjoint representation of ${\rm U}(k)$.
In this way, the low energy effective theory for the D-brane
background is given by the NG modes,
their extension to the gauge multiplets and the gauge field.
%

In addition to the above SSB,
the ${\rm SO}(9,1)$ 
symmetry is broken down to 
${\rm SO}(3,1)\times {\rm SO}(6)$.
However, the NG bosons 
associated with this SSB
do not appear since even an infinitesimal 
rotation of the D3-brane
induces
an infinite change of the field values at space-time infinity.

\subsection{Nambu-Goldstone modes in the IIB matrix model}
\label{sec:NG-IIB}

Let us move on to the discussion of the NG modes in the IIB matrix model
associated with the SSB of
(9+1)-dimensional Poincare symmetry and 
${\cal N}=2$ SUSY.
As in section \ref{sec:local-fields}, we focus on a sufficiently 
small space-time region, 
where 
the space-time can be approximated by 
the direct product of 
$(3+1)$-dimensional Minkowski space 
and 6D small extra dimensions.
Then the ${\rm SO}(9,1)$ symmetry is spontaneously broken 
down to ${\rm SO}(3,1)$, which naively gives rise to
NG bosons corresponding to $\epsilon_{\mu m}$ and $\epsilon_{mn}$ 
in eq.~(\ref{Lorentz transformation}).
However, the NG bosons corresponding to the former do not appear 
since the matrix elements of $\hat{A}_{\mu}$ diverge asymptotically,
and so do the elements of $\delta_{\rm L} \hat{A}_{m}$. 
This is analogous to the situation
with the D-brane background discussed above.
On the other hand, 
the SSB of translational invariance 
in six dimensions yields 6 NG bosons corresponding
to $c_m$ in eq.~(\ref{translation}).
Thus, we obtain 21 NG bosons in total.

Next we discuss the SSB of SUSY.
Here we assume that
ten-dimensional ${\cal N}=2$ 
SUSY is completely broken.
Then, 8 Weyl fermions in four dimensions appear as NG fermions, 
which correspond 
to $\epsilon_1$ and $\epsilon_2$ in eqs.~(\ref{susy1}) and (\ref{susy2}).
Indeed, each of $\epsilon_1$ and $\epsilon_2$, 
which are Majorana-Weyl fermions in ten dimensions, has 16 real components.
In order for the NG fermions corresponding to $\epsilon_1$ to appear,
$[\hat{A}_{M},\hat{A}_{N}]$ in the transformation (\ref{susy1}) 
should be asymptotically finite, which we assume 
in what follows.\footnote{In fact, it is nontrivial whether 
$[\hat{A}_{\mu},\hat{A}_{\nu}]$
are 
asymptotically finite since $\hat{A}_\mu$ represents an expanding universe.
However, we find that this is indeed the case in the physically interesting 
solutions in ref.~\cite{Kim:2012mw}.}

\subsection{Local field theory for the Nambu-Goldstone modes}
\label{sec:lft-NG}

Let us discuss how to derive the local field theory for 
the NG modes. First we substitute (\ref{fluctuation}) 
into the action (\ref{action}).
Since the background is a classical solution, 
there are no terms linear in $\tilde{A}_M$ and $\tilde{\Psi}$.
In order to identify the NG modes, 
we consider the following 
four types of fluctuation:
\begin{align}
&\tilde{A}_{\mu}=0 \ ,\;\;\; 
\tilde{A}_{m} = \phi_{mn}\hat{A}_{n} \ , \;\;\; 
\tilde{\Psi}=0 \ , \label{NGc1}\\
&\tilde{A}_{M} =0 \ , \;\;\; 
\tilde{\Psi} =\frac{i}{2}\Gamma^{MN}[\hat{A}_{M},\hat{A}_{N}]\psi_1
\ , 
\label{NGc2}\\
&\tilde{A}_{M} =0 \ , \;\;\; \tilde{\Psi} =\psi_2 \1 \ , \label{NGc3}\\
&\tilde{A}_{\mu}=0 \ ,\;\;\; 
\tilde{A}_{m} =\rho_{m} \1 \ , \;\;\; \tilde{\Psi} =0 \ , \label{NGc4}
\end{align}
which are associated with (\ref{Lorentz transformation}), 
(\ref{susy1}), (\ref{susy2}) and (\ref{translation}),
respectively. 
Here we have introduced $\phi_{mn}$, $\psi_1$, $\psi_2$ and $\rho_m$
corresponding to $\epsilon_{mn}$, $\epsilon_1$, $\epsilon_2$ and $c_m$, 
which parametrize the spontaneously broken symmetries in the
background. 
Since the action is invariant under 
(\ref{Lorentz transformation}), (\ref{susy1}), 
(\ref{susy2}) and (\ref{translation}),
the quadratic terms with respect to $\phi_{mn}$, $\psi_1$, $\psi_2$
and $\rho_m$ do not appear in the action.
These are the zero modes 
associated with the SSB.

Next we identify the low-lying modes associated with the SSB.
This is done in field theories by making the parameters in the
zero mode fluctuations space-time dependent.
In the matrix model, we make the parameters 
$\phi_{mn}$, $\psi_1$, $\psi_2$ and $\rho_m$ depend 
on the matrix element.
For instance, corresponding to (\ref{NGc1}), we consider
\begin{align}
&(\tilde{A}_{\mu})_{ij}=0 \ ,\;\;(\tilde{A}_{m})_{ij} =
  (\phi_{mn})_{ij}(\hat{A}_{n})_{ij} \ ,
\label{NG1}
\end{align}
and $(\tilde{\Psi})_{ij}=0$,
where 
no sum over $i$ and $j$ is taken.
Obviously, we may set $(\phi_{mn})_{ij}=0$
for $i$ and $j$, which gives $(\hat{A}_{n})_{ij}=0$, 
namely for $\sqrt{(x^{\mu}_i-x^{\mu}_j)^2} \gg \ell$.
When the nonzero components of 
$(\phi_{mn})_{ij}$ are independent of $i$ and $j$, 
(\ref{NG1}) reduces to
(\ref{NGc1}) so that the quadratic terms in the action with respect to
$(\phi_{mn})_{ij}$ disappear.
These facts imply that, for general $(\phi_{mn})_{ij}$,
their quadratic terms coming 
from $\mbox{tr}([\hat{A}_M,\tilde{A}_m]^2)$ \emph{etc.}\
appear in the form of differences between
$(\phi_{mn})_{ij}$ and $(\phi_{mn})_{i'j'}$, where
$x_i$, $x_j$, $x_{i'}$ and $x_{j'}$ are close to each other.
Thus the quadratic terms become local, and 
they are expected to become
the kinetic terms in the continuum limit
due to the (3+1)-dimensional Poincare invariance of the background.
Obviously, the locality is also guaranteed in the higher order terms, 
which represent interactions.

In this way, a local field theory for the NG modes is obtained
from the matrix model.
However, the NG modes interact with each other only through 
derivative couplings, and they decouple at low energy.
Hence it is crucial to introduce the gauge symmetry, which extends 
the NG modes to a set of massless modes. 

The gauge symmetry can be introduced as we discussed 
in section \ref{sec:local-fields}. 
Let us consider the extended background (\ref{extended background}).
In order to discuss the low-lying modes around the background, 
we consider, for instance,
\begin{align}
(\tilde{A}_{\mu})_{ia ,jb}=0 \ , \quad
(\tilde{A}_{m})_{ia, jb} =
  (\phi_{mn})_{ia,jb}(\hat{A}_{n})_{ij}
\end{align}
%
%
%
instead of (\ref{NG1}), and assume that
$ (\phi_{mn})_{ia,jb} = 0$
for $(i,j)$ corresponding to $(\hat{A}_{n})_{ij} = 0$.
(\emph{i.e.}, $\sqrt{(x^{\mu}_i-x^{\mu}_j)^2} \gg \ell$.)
When the nonzero components of $(\phi_{mn})_{ia,jb}$
are independent of $i$ and $j$, 
their quadratic terms do not appear in the action.
Therefore, these modes are guaranteed to be massless at the tree level.
%

The
${\rm U}(k)$ gauge field
is expected to appear
from 
%
$(\tilde{A}_{\mu})_{ia,jb}$. 
The action for $(\phi_{mn})_{ia,jb}$, $(\psi_1)_{ia,jb}$, 
$(\psi_2)_{ia,jb}$, $(\rho_m)_{ia,jb}$ and the gauge field
has the ${\rm U}(k)$ gauge
symmetry if these massless modes transform as
$(\phi_{mn})_{ia,jb} \mapsto u_{ac}^{(i)} (\phi_{mn})_{ic,jd}
  u_{bd} ^{(j)*}$,
\emph{etc.}.\footnote{Strictly speaking, the field $(\phi_{mn})_{ia,jb}$ 
with $i\ne j$ includes the gauge
field as well as the adjoint scalar.}
Since the ${\rm U}(1)$ part of the massless modes is decoupled,
we obtain an ${\rm SU}(k)$ GUT with 21 scalars
and 8 Weyl fermions in the adjoint representation.

\section{Summary and discussions}
\label{sec:concl}

In this paper
we discussed how to 
identify the local fields
corresponding to the massless modes 
that appear at late times 
in the IIB matrix model.
As a concrete example, 
we considered the NG modes associated with
the SSB of Poincare symmetry and SUSY,
which was 
suggested by Monte Carlo results \cite{Kim:2011cr}.
Assuming that a commutative space-time appears at late times,
as suggested by classical solutions \cite{Kim:2012mw}, 
we found that the locality is guaranteed by
the restriction to the NG modes associated with the SSB and their extension.
In general, it is expected that there are other low-lying modes 
than the NG modes.
We have argued that those low-lying modes are also local
if the background
classical solution represents a commutative space time
as defined in section \ref{sec:local-fields}.

The explicit Lagrangian of 
the effective field theory below the Planck scale 
can be derived, in principle,
from the matrix model if we can identify the background
$\hat{A}_M$ dominant at late times 
either by direct Monte Carlo
studies
or by singling out the classical solution, which is 
connected smoothly to Monte Carlo results.
Once we obtain the theory below the Planck scale,
we can use the standard renormalization group analysis
to see whether the SM appears at the electroweak scale.
As we were able to uniquely determine the space-time 
dimensionality \cite{Kim:2011cr}, 
it is possible that the SM is obtained uniquely 
from this top-down approach.

In order to pursue this possibility further, we consider it most 
important to understand the mechanism for the appearance of
chiral fermions.
Some proposals are given already in
refs.~\cite{Aoki:2010gv,Aoki:2012ei,Chatzistavrakidis:2011gs}.
It would be interesting to reconsider these proposals
taking into account the nontrivial dynamics of the Lorentzian model.
As one can see from these works, the structure 
in the extra dimensions is expected to play a crucial role.
From this point of view, it would be also interesting to extend
the work \cite{Kim:2011ts,Kim:2012mw} on the classical solutions
in the Lorentzian model to the cases with
nontrivial structure in the extra dimensions
such as the ones discussed in ref.~\cite{Steinacker:2011wb}.

Another important issue is the breaking of SUSY.
In section \ref{sec:NG-IIB} we have assumed that
the ten-dimensional ${\cal N}=2$ SUSY is completely broken
by the classical solution.
In that case, SUSY is not realized in the particle spectrum,
and all the scalar fields would, generically,
acquire a mass of the order of the GUT scale through radiative
corrections.
This is nothing but the hierarchy problem.
One possibility is that the SM Higgs particle is a composite of
chiral fermions.
Another possibility is that, as opposed to our assumption,
SUSY is only partially broken by
the background above the TeV scale. 

It should be emphasized that
whether the SM appears at the electroweak scale
from the Lorentzian version
of the IIB matrix model
is a completely well-defined question.
We should be able to answer it
in a straightforward manner
if a local field theory is going to appear 
in the way proposed in this paper.
We hope that important clues to the answer 
such as the mechanism for chiral fermions and the SUSY breaking
will be clarified in the near future.

\acknowledgments
We thank H.~Aoki, S.~Iso and S.-W.~Kim
for valuable discussions.
This work is supported in part by Grant-in-Aid
for Scientific Research
(No.\ 20540286, 24540264, and 23244057)
from JSPS.





\end{document}